\documentclass[fleqn,10pt]{wlscirep}
\usepackage[utf8]{inputenc}
\usepackage[T1]{fontenc}
\usepackage{listings}
\definecolor{codegray}{rgb}{0.95,0.95,0.95}
\usepackage{siunitx}
\lstdefinestyle{listing}{
    backgroundcolor=\color{codegray},   
    breakatwhitespace=false,         
    breaklines=true,                 
    captionpos=b,                    
    keepspaces=true,                 
    numbers=left,                    
    numbersep=5pt,                  
    showspaces=false,                
    showstringspaces=false,
    showtabs=false,                  
    tabsize=2
}
\title{Predicting Porosity, Permeability, and Tortuosity of Porous Media from Images by Deep Learning}

\author[1,*]{Krzysztof M. Graczyk}
\author[1]{Maciej Matyka}
\affil[1]{Institute of Theoretical Physics, Faculty of Physics and Astronomy, University of Wroc\l aw, pl. M. Borna 9, 50-204, Wroc\l aw, Poland}

\affil[*]{krzysztof.graczyk@uwr.edu.pl}


\begin{abstract}
Convolutional neural networks (CNN) are utilized to encode the relation between initial configurations of obstacles and three fundamental quantities in porous media: porosity ($\varphi$), permeability $k$, and tortuosity ($T$). The two-dimensional systems with obstacles are considered. The fluid flow through a porous medium is simulated with the lattice Boltzmann method. It is demonstrated that the CNNs are able to predict the porosity, permeability, and tortuosity with good accuracy. With the usage of the CNN models, the relation between $T$ and $\varphi$ has been reproduced and compared with the empirical estimate. The analysis has been performed for the systems with  $\varphi \in (0.37,0.99)$ which covers five orders of magnitude span for permeability $k \in (0.78, 2.1\times 10^5)$ and tortuosity $T \in (1.03,2.74)$.
\end{abstract}
\begin{document}

\flushbottom
\maketitle
\thispagestyle{empty}

\section*{Introduction}

Transport in porous media is ubiquitous: from the neuro-active molecules moving in the brain extracellular space \cite{Rusakov98, Sykova08}, water percolating through granular soils \cite{Nguyen20} till the mass transport in the porous electrodes of the Lithium-ion batteries \cite{Hossain19} used in hand-held electronics. 
The research in porous media concentrates on the understanding connections between two opposite scales: micro-world, that consists of voids and solids, and the macro-scale of porous objects. The macroscopic transport properties of these objects are of the key interest for various industries including healthcare \cite{Suen19} and mining \cite{Huaqing18}.

Macroscopic properties of the porous medium rely on the microscopic structure of interconnected pore space. The shape and the complexity of pores depend on the type of medium. Indeed, pores can be elongated and interwoven showing a high porosity and anisotropy like in fibrous media \cite{Koponen98,Shou11}. On the other hand, medium can be of low porosity with a tight network of twisted channels. For instance, in rocks and shales, as a result of erosion or cracks, large fissures can be intertwined in various ways~\cite{Huaqing18}.

Porosity ($\varphi$), permeability ($k$), and tortuosity ($T$) are the three parameters that play an important role in description and understanding of the transport through a porous medium. The porosity is the fundamental number that describes the fraction of voids in the medium. The permeability reflects the ability of the medium to transport the fluid. Eventually, the tortuosity characterizes paths of particles being transported through the medium\cite{Backeberg17,Niya18}. 
The porosity, permeability, and tortuosity are related by the Carman-Kozeny~\cite{Koponen97} law, namely:
\begin{equation}
k=\frac{\varphi^3}{cT^2S^2},
\label{Eq:Karman}
\end{equation}
where $S$ is the specific surface area of pores and $c$ is the shape factor. The tortuosity is defined as the elongation of the flow paths in the pore space \cite{Matyka08,Ghanbarian13}:
\begin{equation}
    T=\frac{L_\mathrm{eff}}{L},
\end{equation}
where $L_\mathrm{eff}$ is an effective path length of particles in pore space and $L$ is a sample length ($L<L_{eff}$ and thus, $T>1$). 
The $T$ is calculated from the velocity field obtained either experimentally or numerically. There are experimental techniques for particle path imaging in porous medium, like particle image velocimetry~\cite{Morad09}. However, they have limitations due to the types of porous medium. There are two popular numerical approaches for estimation of $T$. In the first, particles paths are generated by integration of the equation of motion of fluid particles~\cite{Koponen97,Matyka08}. In the other, used in this work, $T$ is calculated directly from the velocity field by averaging its components\cite{Koponen97,Duda11}.  

Calculating the flow in the real porous medium with the complicated structure of pores requires either a special procedure for grid generation in the standard Navier-Stokes solvers \cite{Boccardo20} or usage of a kind of mesoscopic lattice gas-based methods \cite{Bakhshian19}.
Nevertheless the type of solver used for the computation, the simulation procedure is time and computer resource consuming. Thus, we propose the convolutional neural networks (CNN) based approach to simplify and speed up the process of computing the basic properties of a porous medium.

The deep learning  (DL) \cite{LeCun2015} is a part of the machine learning  and the artificial intelligence methods. It allows to analyze or describe complex data. It is useful in optimizing the complicated numerical models. Eventually, it is a common practice to use the DL in problems not described analytically. The DL finds successful applications to real-life problems \cite{1986Natur.323..533R} like automatic speech recognition, visual-image recognition, natural language processing, medical image analysis, and others. The DL has become an important tool in science as well.

Indeed, recently, the number of applications of the DL methods to problems in physics and material science grows exponentially. A recent review of the applications in physics can be found in \cite{MEHTA20191}. One of the standard applications is to use the machine learning models to analyze the experimental data~\cite{Graczyk:2014lba}. The DL networks  are utilized to study the phase structure of the quantum and classical matters~\cite{carrasquilla2020machine}. The DL is  applied to solve the ordinary differential equations \cite{LAGARIS19971,zbay2019poisson,pannekoucke2020pdenetgen}.  There are applications of the DL approaches to the problems of the fluid dynamics. Here, the main idea is to obtain the relations between the system represented by the picture and the physical quantities, like velocities, pressure etc.

The DL is one of the tools considered in the investigation of the transport in a porous medium. For instance, the neural networks are used to obtain porous material parameters from the wave propagation~\cite{Lahivaara18}. Permeability is evaluated using the multivariate structural regression \cite{Andrew20} or directly from the images \cite{Wu2018}. The fluid flow field predictions using convolutional neural networks based on the sphere packing input data are given in Ref.~\cite{Santos2020}. The advantages of the usage of the machine learning techniques over physical models in computing permeability of cemented sandstones is shown in \cite{Male20}. The predictions of the tortuosity for unsaturated flows in porous media based on the scanning electron microscopy images have been performed \cite{Zhang20}. Eventually, diffusive transport properties have been investigated via convolutional neural networks working on images of the media geometry only~\cite{Wu2019}.

Our goal is to find the dependence between the configurations of obstacles, represented by the picture, and porosity, permeability, and tortuosity. The relation will be encoded by the convolutional neural network which for the input takes the binary picture with obstacles and it gives as an output the vector $(\varphi, k, T)$. 
To automate, simplify, and optimize the process we use the state of the art CNN networks and DL techniques combined with the geometry input and fluid solver based outputs. Namely, we consider a large number of synthetic, random, porous samples with controllable porosity to train the networks. The training data set contains the porous media geometries with corresponding numerical values of porosity, permeability and tortuosity. The two latter variables are obtained from the flow simulations done with the lattice Boltzmann solver. 

As the result of the analysis we obtain two CNN models which predict the porosity, permeability, and tortuosity with good accuracy. The relative difference between the predictions and the 'true' values, for our the best model, does not exceed $6 \%$. 
Eventually,  we generate a blind set of geometry data for which we predict the tortuosity and porosity. The obtained $T(\varphi)$ dependence is in qualitative agreement with the empirical fits.

The paper is organized as follows: in Sec.~\ref{sec:fluid-flow} we introduce the lattice Boltzmann method. Sec.~\ref{sec:nn} describes the CNN approach and method of data generation. Sec.~\ref{sec:results} contains the discussion of the numerical results and summary.

\section{Lattice Boltzmann method}
\label{sec:fluid-flow}

The data set consists of the rectangular pictures of the configuration of obstacles and corresponding quantities (labels): $\varphi$, $k$, and $T$, which are calculated from the lattice Boltzmann method (LBM) flow simulations.

The LBM has confirmed its capability to solve complex flows in complicated porous geometries~\cite{Succi01}. It is based on the density distribution function, which is transported according to the discrete Boltzmann equation. In the single relaxation time approximation the transport equation reads:
\begin{equation}
\frac{\partial f}{\partial t} + \vec{v}\cdot\nabla f = - \frac{1}{\tau}(f-g),
\end{equation}
where $f(\vec{r},t)$ is the density distribution function, $\vec{v}$ is the macroscopic velocity, $\tau$ is the relaxation time and $g$ is the Maxwell-Boltzmann distribution function at given velocity and density (see e.g. \cite{He97,Duda11}). 

\begin{figure}
    \centering
\includegraphics[width=0.3\columnwidth]{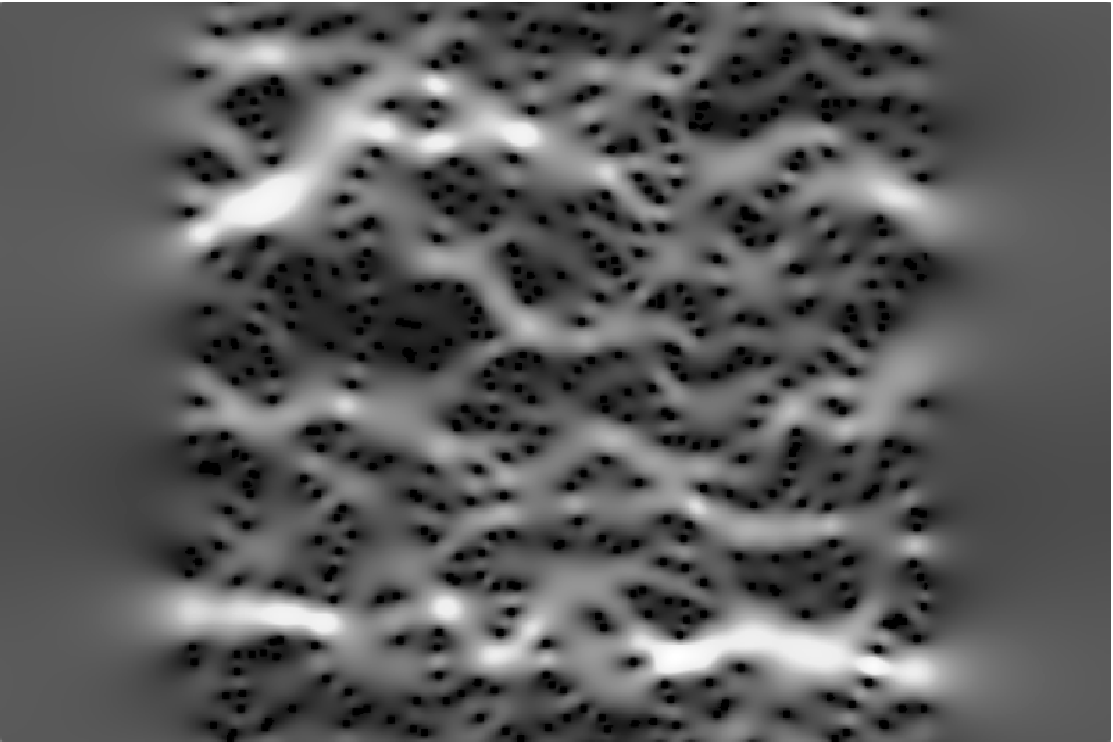}
\includegraphics[width=0.3\columnwidth]{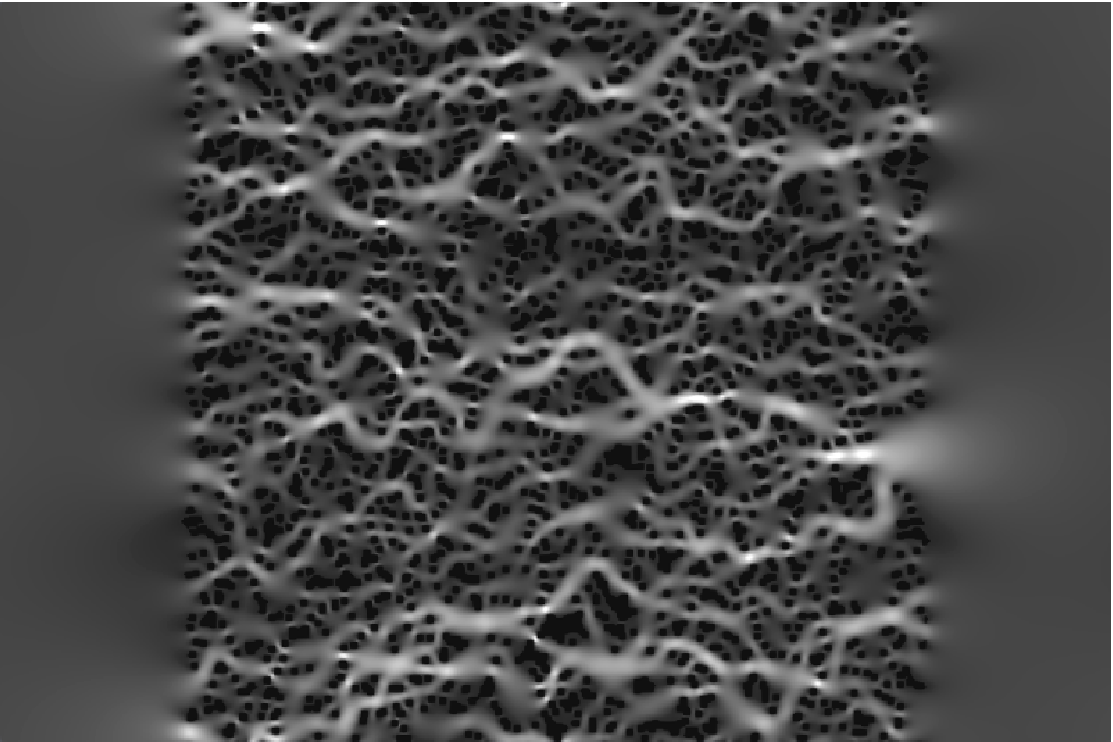}
\includegraphics[width=0.3\columnwidth]{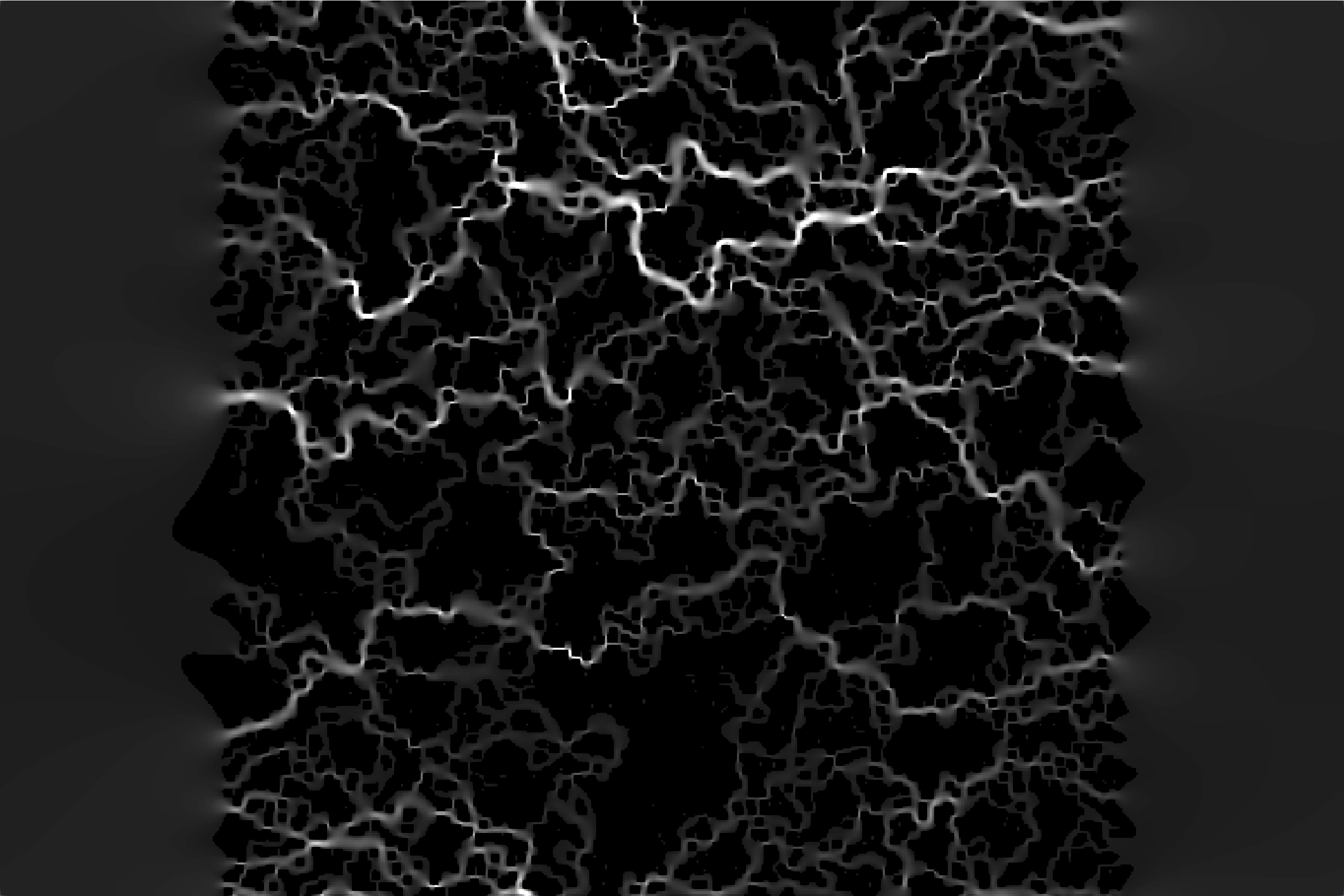}
    \caption{An exemplary pore-scale velocity magnitude (the brighter the color the larger velocity) in the fluid flow calculated using the LBM based on configurations from Fig.~\ref{fig:examples}. Contrast and brightness were adjusted to better visualize the structure of emerged flow paths.
    \label{fig:lbmresults}}
\end{figure}

The LBM solver provides with an information about the time evolution of the density function from which the velocity field is obtained. We use the pore-scale approach (the obstacles are completely solid and pores are completely permeable), thus, the velocity field is solved at pore space. 
Each simulation starts with the zero velocity condition. The sample is exposed to an external gravity force that pushes the flow. Notice that we keep the number of iterations larger than $10000$ and less than $1000000$. 
For the steady-state condition we take the sum of the relative change of the velocity field in the pore space:
\begin{equation}
    c = \sum_{i,j} \frac{(u_{i,j}-u_{i,j}')^2}{u_{i,j}^2}
\end{equation}
where $u$ is the local velocity at the current time step (taken at nodes $i$,$j$) and $u'$ is the velocity at time step $500$ time steps earlier. 
The local change of the velocity is monitored to verify if the steady-state is achieved. 
The exemplary velocity field resulting from the above procedure is shown in Fig.~\ref{fig:lbmresults}. 

\section{Deep learning approach}
\label{sec:nn}

\subsection{Convolutional neural networks}

The convolutional neural networks are used to encode the dependence between the initial configuration of obstacles and the porosity, tortuosity as well as permeability. 
The CNN is a type of deep neural network designed to analyze the multi-channel pictures. It has been successfully applied for the classification \cite{Lecun98} and the non-linear regression problems \cite{nair2010rectified}. 

The CNN consists of sequence of convolutional blocks and fully connected layers. In the simplest scenario a single block contains a convolutional layer with a number of kernels. Usually, the pooling layer follows on the convolutional layer while in the hidden layer the rectified linear units (ReLU) \cite{nair2010rectified} are considered as the activation functions. For the theoretical foundations of the convolutional networks see Ref.~\cite{Goodfellow-et-al-2016}.

A kernel extracts single feature of the input. The first  layer of CNN collects the simplest objects (features) such as edges, corners, etc. The next layers relate extracted features. The role of the pooling layer is to amplify the signal from the features as well as to reduce the size of the input. Usually, a sequence of fully connected layers follows on the section of the convolutional blocks. 

The CNNs considered in this work contain the batch normalization layers \cite{ioffe2015batch}. This type of the layer has been proposed to maintain the proper normalization of the inputs. It has been shown that having the batch normalization layers allows one to improve the network performance on the validation set. Moreover, it is a method of regularization, which is alternative to the  dropout technique \cite{JMLR:v15:srivastava14a}. 
\begin{figure}
    \centering
 \includegraphics[width=0.3\columnwidth]{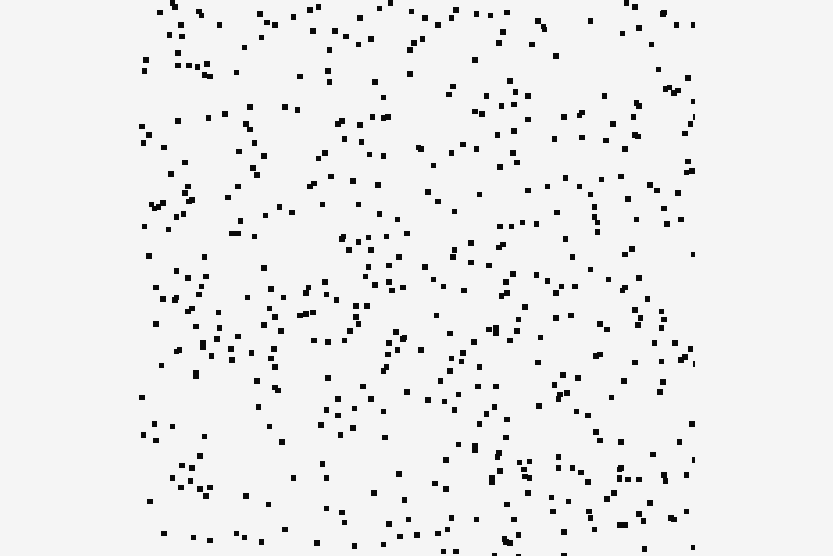}
 \includegraphics[width=0.3\columnwidth]{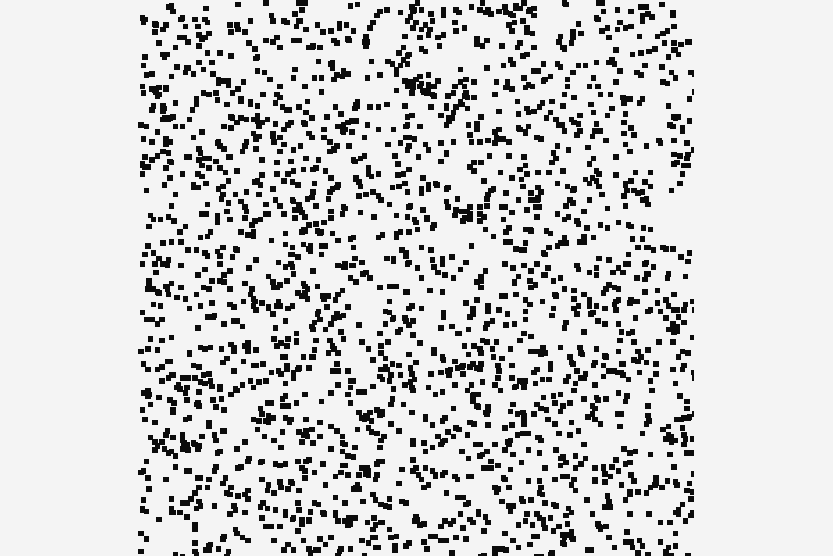}
 \includegraphics[width=0.3\columnwidth]{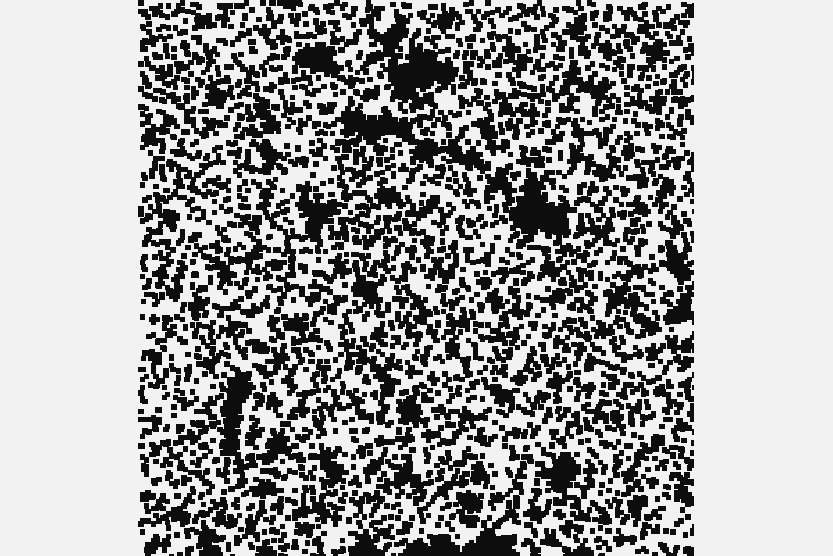}
    \caption{Exemplary random porous samples at porosity $\varphi=0.95, 0.8$ and $0.5$ (from left to right), size $800\times 400$. Black blocks represent obstacles for the flow and the interconnected light gray area is the pore space filled by fluid. Black clusters visible for $\varphi=0.5$ are the effect of the filling-gap algorithm used for prepossessing of the data after generating with random deposition procedure. The gaps are not accessible for the fluid and, thus, we fill them before they are provided to the fluid solver and neural network.}
    \label{fig:examples}
\end{figure}

\subsection{Data}

\begin{figure}
    \centering
    \includegraphics[width=\textwidth]{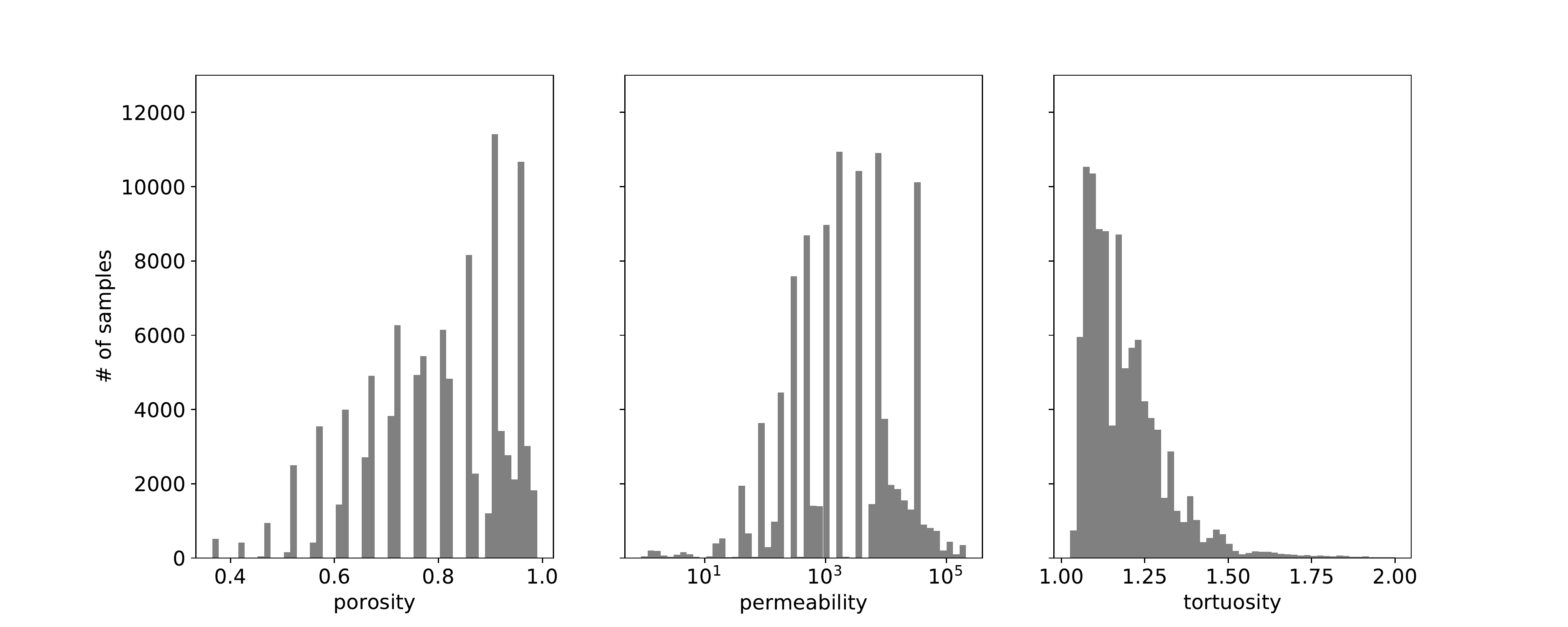}
    \caption{The distribution of the generated samples (training and validation data sets together) with respect to porosity, permeability, and tortuosity. \label{Fig:distribution-all}}
\end{figure}
Random deposition (RD) adsorption models are popular for generating porous structures for numerical solvers~\cite{Chueh14, Li18}.  We use the RD model of a porous medium. The samples are build of overlapping quad solids laying on the two-dimensional surface.

A total number of $100, 000$ one-channel figures with obstacles and predicted values of the porosity, tortuosity, and the permeability have been prepared. A given figure, see Fig.~\ref{fig:examples}, is the binary picture of the size $800\times 400$. It includes two vertical margins of the width $200$ (pixels) each. The margins are kept to reduce the influence of boundaries on the calculations of porosity, tortuosity, and permeability. But from the point of view of the CNN analysis, they do not contain the information. Hence, in the training and inference process the picture without margins is taken as an input for the network. Effectively, the input has the size $400\times 400$.  

Each sample (figure) consists of $400 \times 400$ nodes which are either free or blocked for the fluid flow. The periodic boundary condition is set at sides with margins, whereas two remaining sides are considered no-slip boundary condition. To generate medium at given porosity we start with an empty system and systematically add quad obstacles at random positions in the porous region. The margins are excluded from this deposition procedure. We do not consider blocking of sites, thus, obstacles may overlap freely. Each time an obstacle is placed, we update current porosity and stop if the desired value is reached. In the next step of the generation of figures, we use the simple flood-filling algorithm to eliminate all the non-physical islands from each sample\footnote{Island is the pore volume completely immersed in solids.}. The exemplary porous samples at four different porosity values are shown in Fig.~\ref{fig:examples}.
Having the binary images of the pore space the LBM solver calculates the velocity distribution from which the quantities of interest: porosity, tortuosity, and permeability are obtained. 

The success of the training of the network is determined by the quality of the data, its representative ability. 
The data should uniformly cover the space of label parameters. The ranges of porosity, permeability, and tortuosity, for the systems discussed in our analysis, are given in Table~\ref{Tab:ranges}.  
In the initial procedure, the porosity of samples was chosen from the uniform distribution. However, some of the samples were not permeable and, thus, we notice the skewed character of porosity distribution (see Fig. \ref{Fig:distribution-all}, left). As a consequence, the tortuosity distribution is skewed as well with most of the systems at $T<1.5$. We found a few systems with $T>2.0$ at low permeability but they were rejected from the analysis as the predictions of the LBM solver are uncertain in this range.
Moreover, in order to improve the learning process for higher porosity values, we generated additional samples with porosity $\varphi>0.85$.
\begin{table}
\begin{center}
\begin{tabular}{|c|c|c|c|}
\hline\hline
  & Porosity & Permeability & Tortuosity \\
\hline
min & $0.37$ & $0.78$ &  $1.03$         \\
\hline
max & $0.99$ & $2.1\times 10^5$ & $2.74$ \\
\hline
\end{tabular}
\end{center}
\caption{The ranges of the porosity, permeability, and tortuosity obtained for the ensemble of $100,000$ samples of the systems generated within the fluid flow simulation procedure. \label{Tab:ranges}}
\end{table}
As a result, an important fraction of the relevant labels is included in the tail of the distributions. Training the network with such distributed data will lead to the model with excellent performance on the systems with numbers characteristic for the distribution peaks and low predictive ability for the systems which are outside of the peaks. We partially solve this  problem by considering the re-weighted distribution of samples, namely,
\begin{itemize}
    \item the data are binned in the two-dimensional histogram in permeability and tortuosity;
    \item for every bin we calculate the weight given by the ratio of the total number of samples to the number of samples in a given bin; 
    \item the ratios form the the re-weighted distribution of samples;
    \item during the training, labelled figures in the mini-batch, are sampled from the re-weighted distribution. 
\end{itemize}

Another standard procedure, adapted by us, is the transformation of labels, so that, they have the values on the neighborhood of zero. 
\begin{eqnarray}
    \varphi & \to & \varphi/\varphi_{s} - \overline{\varphi} \\
    T & \to & T/T_{s} - \overline{T}.
\end{eqnarray}
In the case of the permeability, instead of $k$, we consider $\log$ of $k$ and do re-scaling:
\begin{equation}
    k \to \log(k)/\log(k_s) - \log(\overline{k}).
\end{equation}
In the final analyses the following settings are used: $\varphi_{s}=1.0$, $\overline{\varphi}=0.5$, $T_{s}=2.8$, $\overline{T}=0.5$ as well as $\log(k_s)=12.3$, $\log(\overline{k}) = 0.5$.

\begin{figure}
    \centering
\includegraphics[width=0.9\textwidth]{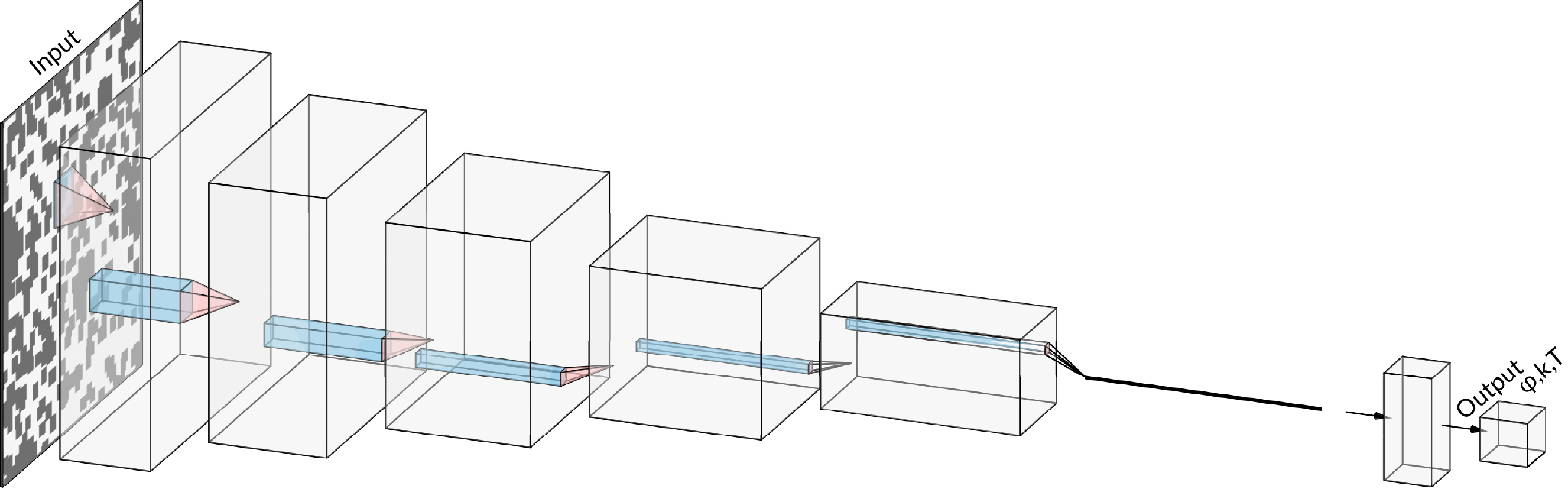}
\caption{Network $net_{A}$ architecture. It contains six convolutional blocks and two (including output) fully connected layers. 
Each convolutional block contains the max pooling layer. The first three convolutional blocks consist of batch normalization layers. Graph was drawn using \cite{AlexImage}.}.
    \label{Fig:architecture_net200}
\end{figure}

\begin{table}[ht]
    \begin{center}
    \begin{tabular}{|c|c|c|c|}
                          \hline
    \multicolumn{4}{|c|}{ $net_A$}  \\ 
    \hline
                          &  $\varphi$ & $k$ & $T$ \\
                          \hline
    \multicolumn{4}{|c|}{training set}  \\ 
    \hline
        $\overline{R}$              &     \num{-5.27e-04}     &   \num{-8.63e-04}         &   \num{-1.65e-04}         \\
        $\sqrt{Var(R)}$ &     \num{9.40e-03}      &   \num{7.54e-02}           &  \num{2.47e-03}          \\
        \hline
        \multicolumn{4}{|c|}{validation set}  \\
        \hline
        $\overline{R} $   &     \num{-5.83e-04}     &   \num{-8.88e-04}         &   \num{-1.37e-04}         \\
        $\sqrt{Var(R)}$ &     \num{9.33e-03}      &   \num{5.55e-02}           &  \num{2.45e-03}          \\
        \hline
    \end{tabular}
        \begin{tabular}{|c|c|c|c|}
    \hline
         \multicolumn{4}{|c|}{ $net_B$}  \\ 
     \hline
                          &  $\varphi$ & $k$ & $T$ \\
     \hline
    \multicolumn{4}{|c|}{training set}  \\ 
    \hline                     
        $\overline{R}$              &     \num{-2.64e-04}     &   \num{-1.91e-04}         &   \num{2.81e-06}         \\
        $\sqrt{Var(R)}$ &     \num{7.44e-03}      &   \num{3.35e-02}           &  \num{2.01e-03}          \\
        \hline
         \multicolumn{4}{|c|}{validation set}  \\
         \hline
        mean              &     \num{-3.64e-04}     &   \num{-1.48e-03}         &   \num{-1.70e-05}         \\
        $\sqrt{Var(R)}$ &     \num{7.31e-03}      &   \num{6.28e-01}           &  \num{2.07e-03}          \\
        \hline
    \end{tabular}
    \end{center}
    \caption{The mean of $\overline{R}$ and $\sqrt{Var(R)}$  computed from the predictions of networks $net_A$ and $net_B$. $R$ is defined by Eq.~\ref{Eq:ratio}.} 
    \label{tab:results_for_A_B}
\end{table}

\subsection{Network architecture}

In order to define the CNN architecture we introduce the blocks: 
\begin{itemize}
\item $C(N, K, S, P, act)$ convolutional layer with: $N$ kernels of the size $K\times K$ with the stride $S$, padding $P$ and $act$ - activation function;
\item $MP(K=2) \equiv MP()$ max pooling layer of the size $K\times K$;
\item $B()$ - batch normalization layer;
\item $F(M, act)$ fully connected layer with activation function $act$.
\end{itemize}

The code has been implemented using the PyTorch library~\cite{NEURIPS2019_9015}. We distinguish two types of
 the analyses: 
\begin{itemize}
    \item[(A)] the input figures are re-sized to $200\times 200$;
    \item[(B)] the input figures are original size $400 \times 400$.
\end{itemize}    
In the first case, the training of the network is faster, however, a fraction of the information hidden in the figures might be lost. In the other, 
the training is slower but full information is used in the analysis.

For the analysis (A) we consider a network $net_A$. It contains six convolutional blocks and two (including output) fully connected layer (see Fig.~\ref{Fig:architecture_net200}), namely: 
\begin{eqnarray}
    input &\to & C(10, 10, 1, 0, \mathrm{ReLU})\cdot B() 
          \cdot MP() \nonumber \\
          & \to &  C(20, 7, 1, 0, \mathrm{ReLU})\cdot B() 
           \cdot MP() \nonumber \\
          & \to &  C(40, 5, 1, 0, \mathrm{ReLU})\cdot B() \cdot MP()
          \nonumber  \\
          & \to &  C(80, 3, 1, 0, \mathrm{ReLU})\cdot MP()
          \nonumber \\
          & \to &  C(160, 2, 1, 0, \mathrm{ReLU})\cdot MP()
          \nonumber  \\
          & \to &  C(400, 2, 1, 0, \mathrm{ReLU})\cdot MP()
          \nonumber  \\
          & \to & F (10,\mathrm{\tanh}) \nonumber\\
                    \label{nn_scheme_200}
           & \to &  F (3,id) = output,
\end{eqnarray}
where $id$ is identity map and $\tanh$ refers to hyperbolic tangent.

The network, $net_B$, used in the analysis (B), contains seven convolutional blocks and two (including output) fully connected layer, namely: 
\begin{eqnarray}
    input &\to & C(10, 5, 1, 2, \mathrm{ReLU}) \cdot MP() \nonumber  \\
          &\to & C(10, 10, 1, 0, \mathrm{ReLU})\cdot B() 
          \cdot MP() 
          \nonumber  \\
          & \to &  C(20, 7, 1, 0, \mathrm{ReLU})\cdot B() 
           \cdot MP() 
           \nonumber \\
          & \to &  C(40, 5, 1, 0, \mathrm{ReLU})\cdot B() \cdot MP()
          \nonumber  \\
          & \to &  C(80, 3, 1, 0, \mathrm{ReLU})\cdot MP()
          \nonumber  \\
          & \to &  C(160, 2, 1, 0, \mathrm{ReLU})\cdot MP()
          \nonumber  \\
          & \to &  C(400, 2, 1, 0, \mathrm{ReLU})\cdot MP()
          \nonumber  \\
          & \to & F (10,\mathrm{\tanh})  \nonumber\\
                    \label{nn_scheme_400}
         &   \to & F (3,id) = output
\end{eqnarray}
%http://alexlenail.me/NN-SVG/LeNet.html
Notice that the first three CNN blocks in $net_A$ and three subsequent blocks (starting from second) in $net_B$ contain the batch normalization layers. 

\begin{figure}[p]
    \centering
    \includegraphics[width=\textwidth]{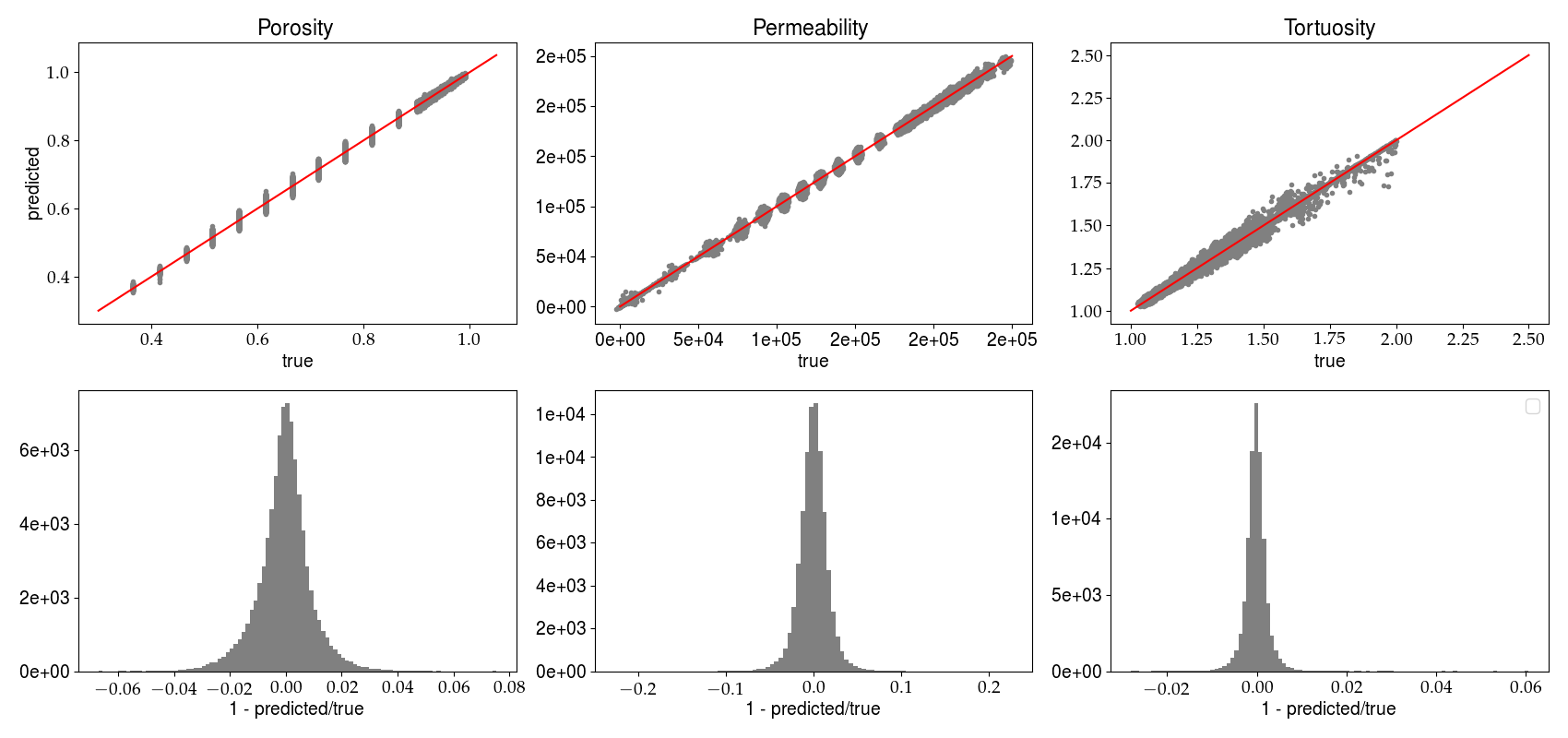}
    \caption{Predictions of porosity, permeability, and tortuosity by CNN versus `true` data (upper row). In the bottom row, the histograms of $R$, see Eq. \ref{Eq:ratio}, are plotted. The results are obtained for analysis (A) and for the training data set. Solid line, in the top row, represents $predicted=true$ equality.
}
    \label{fig:cnnresults_validnet_trainset_200}
\end{figure}
\begin{figure}[p]
    \centering
    \includegraphics[width=\textwidth]{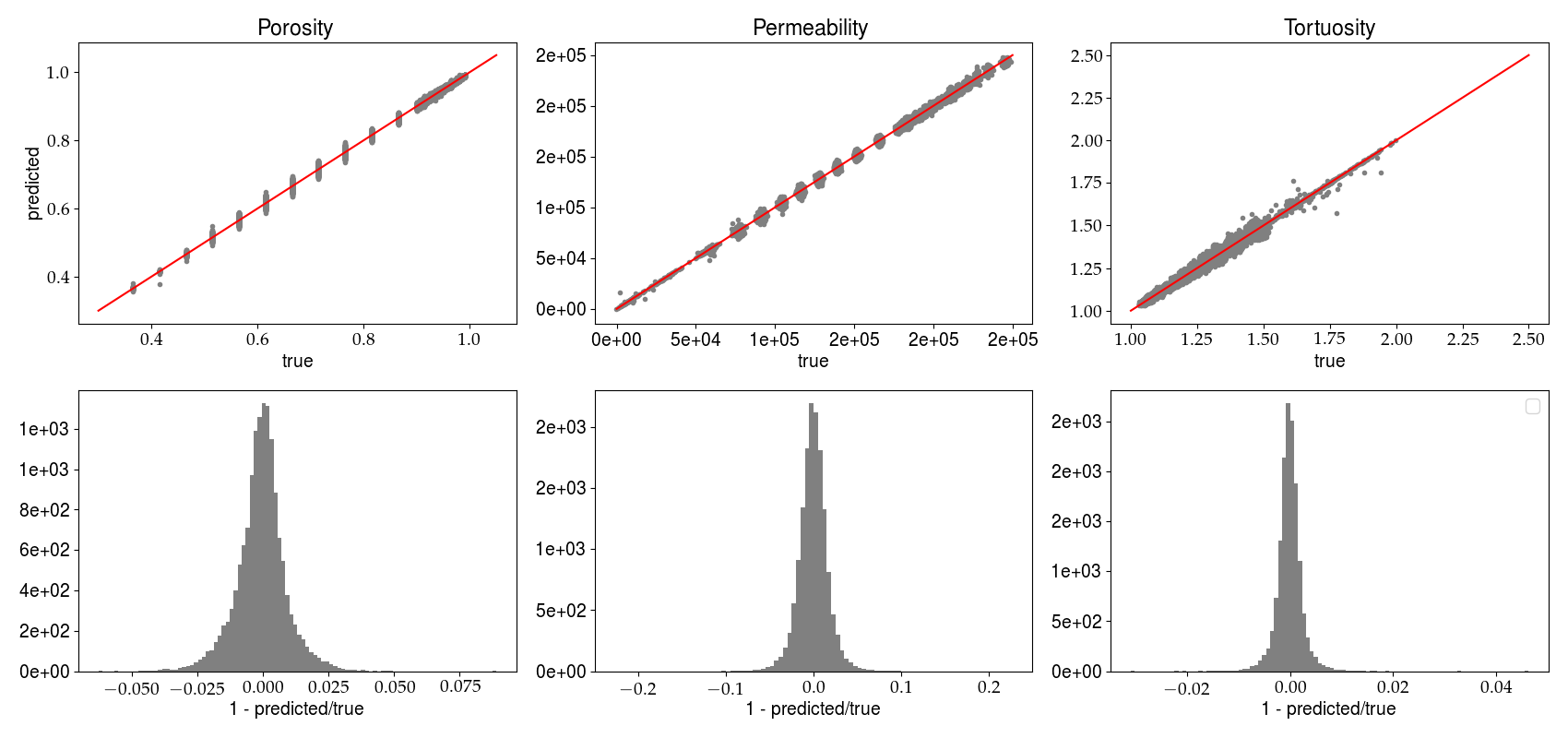}
    \caption{Caption the same as in Fig.~\ref{fig:cnnresults_validnet_trainset_200} but the predictions are made for  the validation data set. 
}
    \label{fig:cnnresults_validnet_testset_200}
\end{figure}
\begin{figure}[p]
    \centering
    \includegraphics[width=\textwidth]{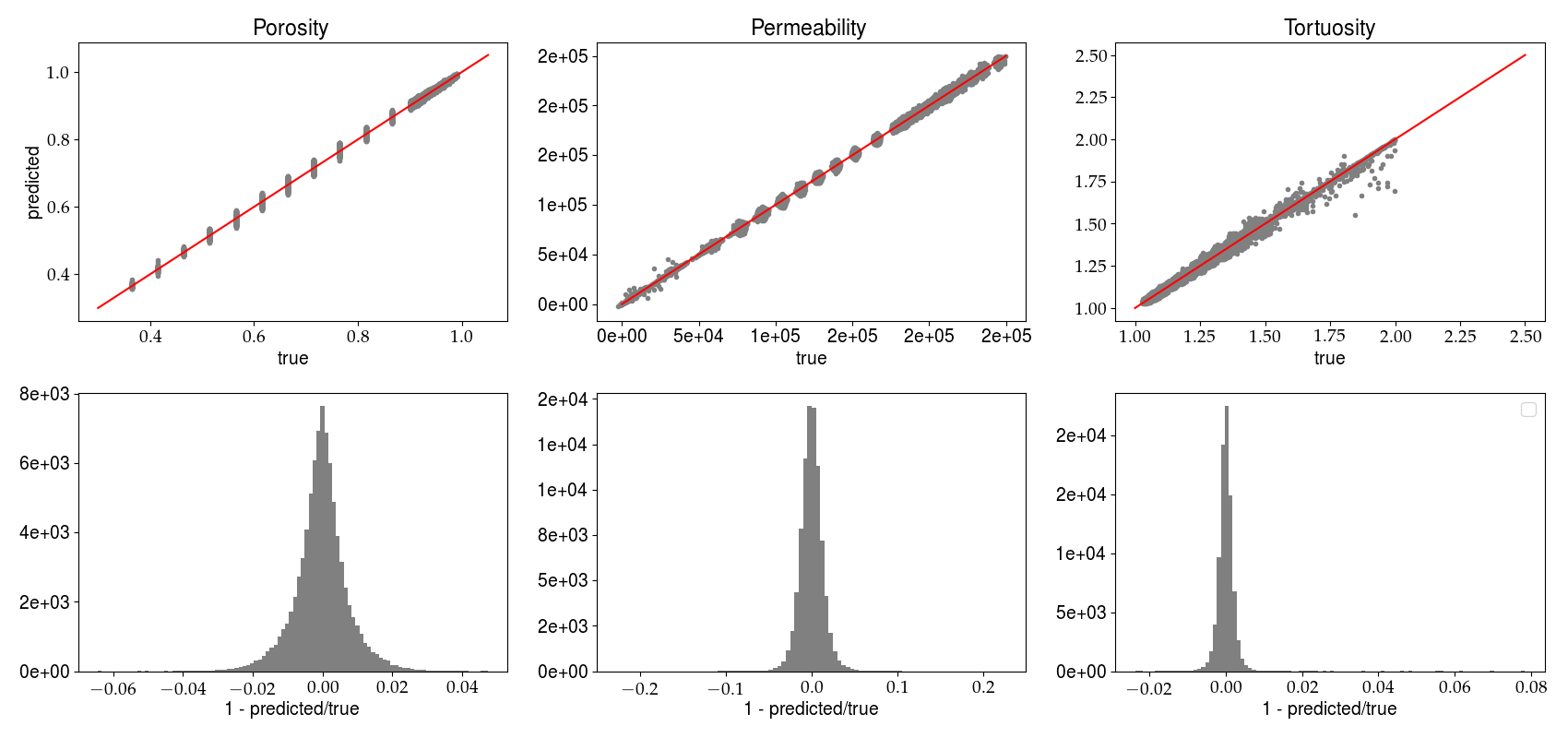}
    \caption{Predictions of porosity, permeability, and tortuosity by CNN versus `true` data (upper row). In the bottom row, the histograms of $R$, see Eq. \ref{Eq:ratio}, are plotted. The results are obtained for analysis (B) and for the training data set.  Solid line, in the top row, represents $predicted=true$ equality.
}
    \label{fig:cnnresults_validnet_trainset_400}
\end{figure}
\begin{figure}[p]
    \centering
    \includegraphics[width=\textwidth]{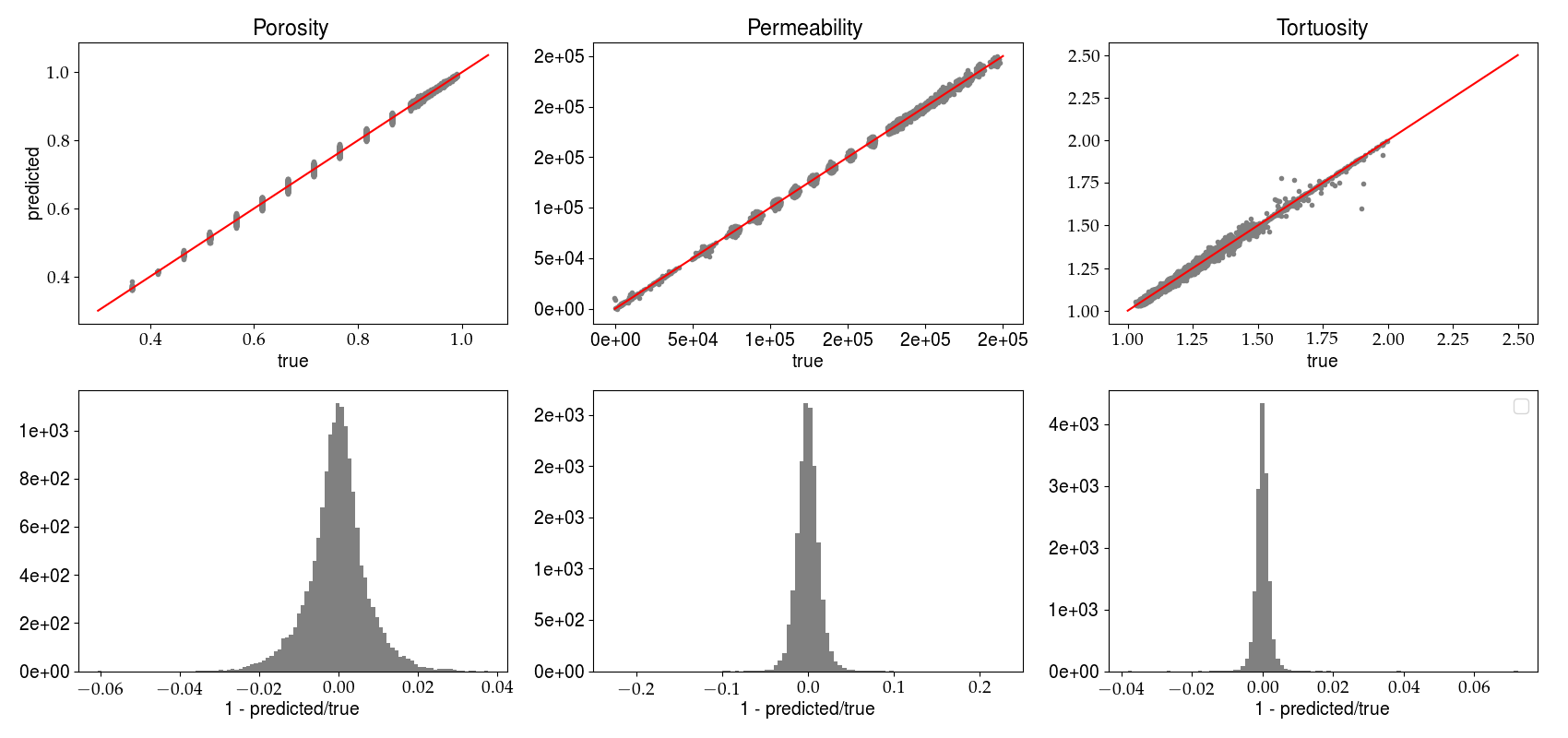}
    \caption{Caption the same as in Fig.~\ref{fig:cnnresults_validnet_trainset_400} but the predictions are made for  the validation data set. 
}
\label{fig:cnnresults_validnet_testset_400}
\end{figure}

\begin{figure}[ht]
    \centering
    \includegraphics[width=0.75\columnwidth]{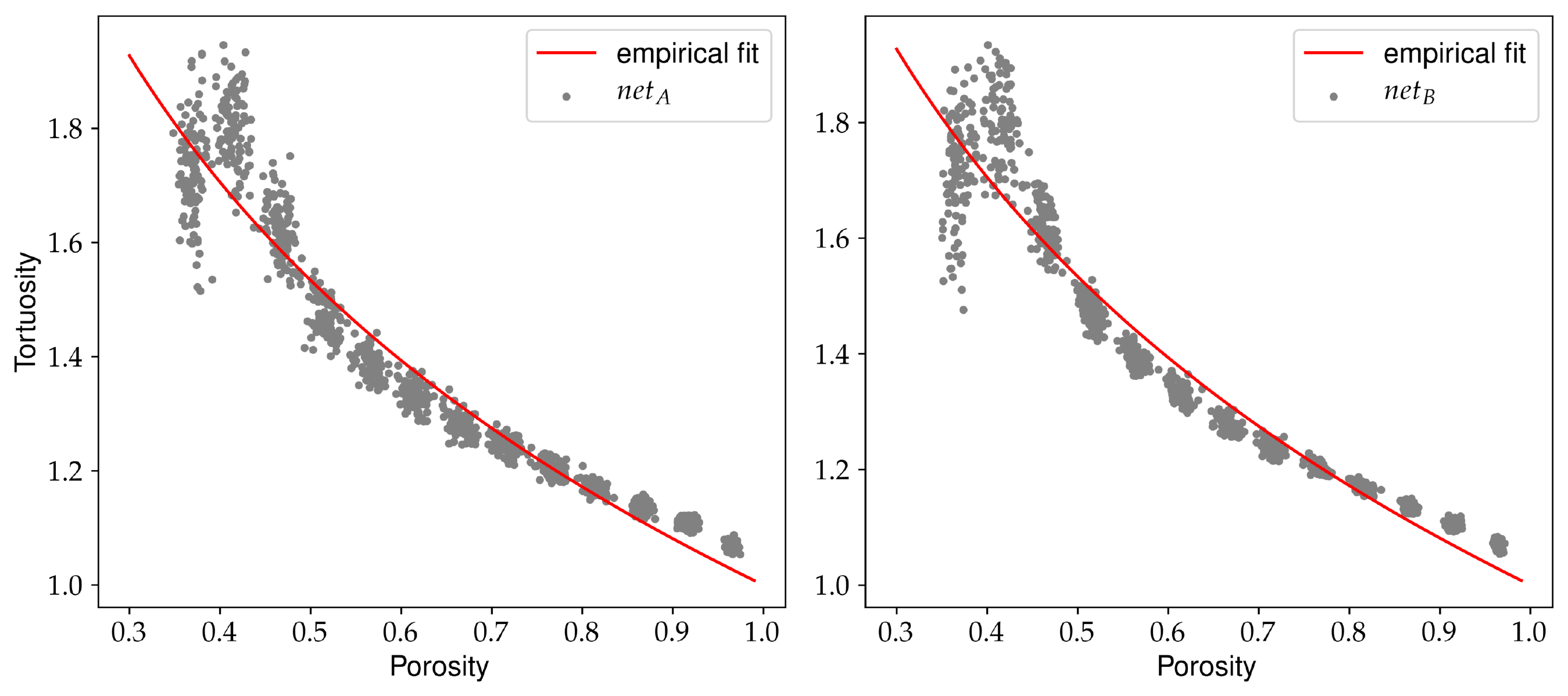}
    \caption{Tortuosity versus porosity: solid line represents empirical relation $T(\varphi)=1-0.77 \log(\varphi)$ obtained within classical fluid flow approach \cite{Matyka08}. The points are obtained from the unlabelled data set. 
    \label{fig:porosity_vs_tortuosity}}
\end{figure}
\subsection{Training}

The prepared data set has been split into the training ($85\%$ of the total number) and validation ($15 \%$ of the total number) data sets. After the cut in the tortuosity, we have  84,917 and 14,986 samples in training and validation data sets, respectively.

In the pre-analysis, we considered two types of loss functions, namely, the mean absolute error (MAE) and mean square error (MSE).  The MAE is more sensitive to the outliers. However, our preliminary experiments showed that optimization of the MSE leads to the networks with better performance on the validation set than the models optimized with the MAE. Therefore,  we report the results obtained for the models for which the MSE loss is considered.
During each epoch step we calculate the error on the validation data set. The model with the smallest error value is saved. 

The stochastic gradient descent (SGD) algorithm, in the mini-batch version, has been utilized for the optimization \cite{Bottou2012,pmlr-v28-sutskever13}. The mini-batch contains $250$ (analysis (A)) and $65$ (analysis (B) samples. The SGD is one of the simplest learning algorithms but it is known that it naturally regularize the model and the obtained networks tend to have better performance on the validation data set~\cite{doi:10.1162/neco.1995.7.1.108,keskar2016largebatch,1986Natur.323..533R}.

\section{Results and summary}
\label{sec:results}

One of the difficulties in the DL analyses is a proper choice of the hyperparameters, such as: the size of the mini-batch, the   optimization algorithm parameters, number of the CNN blocks, the size of the kernels, etc. We experimented with various configurations of the hyperparameters. Eventually, we established the final  settings, namely, the SGD algorithm is run with the momentum $0.9$ and the initial value of the learning rate $0.1$. The latter parameter is reduced by $10\%$ every $50$ epochs of the training. The size of the mini-batches as well as network architectures are reported in the previous section. 
The both network architecture are designed so that the number of filters in the CNN increases with the depth of the network but the size of the inputs is reduced. The kernels in the first layer of the $net_B$ are large ($K=10$) to capture the long-distance correlations. In the subsequent layers the kernels have smaller sizes. The $n_B$ is the network $net_A$ with additional input layer. In the case of the $net_A$ instead of the layer with $K=10$ the input is re-sized. In both network schemes (A) and (B) the output of the last CNN layer is a vector of the length $400$.

The best model is the one with the smallest error on the validation set. Figs.~\ref{fig:cnnresults_validnet_trainset_200} and \ref{fig:cnnresults_validnet_testset_200} present the results for  model $net_A$, whereas Figs.~\ref{fig:cnnresults_validnet_trainset_400}
and \ref{fig:cnnresults_validnet_testset_400} show the results for the analysis (B). 
In each figure, the predictions of porosity, permeability, and tortuosity versus the 'true' (as obtained from the LBM solver) values are plotted. Additionally, each figure contains the histograms of ratio
\begin{equation}
\label{Eq:ratio}
R = 1 - \frac{predicted}{true},
\end{equation}
where  $R$ was computed for porosity, permeability and tortuosity. More quantitative description of the histograms is given in  Table \ref{tab:results_for_A_B}, in which  the mean ($\overline{R}$) and variance $\sqrt{Var(R)}$ computed for all presented histograms are given. 
Notice that we present the network predictions for the training and validation data sets.

The agreement between the networks predictions and 'true' values seems to be pretty good. The performance of the both networks is comparable. However,  $net_A$ with the resized input seems to work more efficiently  on the validation data set.
It both analyses it was relatively easy to fit the porosity. The difficulties appeared in modeling the permeability and  tortuosity. Indeed, it was hard to get an accurate model predictions for low permeability and the tortuosity $T>1.75$. The re-weighting procedure, described above, allowed to partially solve the problem. Another improvement came from the inclusion of the batch normalization layers.

In the analysis (A) for a large fraction of samples the relative difference between the network response and the 'true' value is smaller than $6\%$. The most accurate predictions are obtained for porosity and tortuosity. Indeed, the difference between 'true' and predicted is smaller than $1\%$. Model $net_B$ predicts the porosity and tortuosity with similar accuracy as the model $net_A$ but the the predictions of the permeability is rather uncertain in this case.  

Both data sets, training and validation are used in the optimization process.  Hence to examine the quality of obtained models we generated the third set of the data. It contains $1,300$ unlabelled samples. This data set is used to reconstruct the dependence between porosity and tortuosity, see Fig. \ref{fig:porosity_vs_tortuosity}). The obtained $T(\varphi)$ relation   agrees qualitatively with the empirical fit  from the analysis of the fluid flow ~\cite{Matyka08}. Indeed, 
tortuosity grows with $\varphi \rightarrow \varphi_c$, where $\varphi_c$ is the percolation threshold ($\varphi\approx 0.4$ for overlapping quads model \cite{Koza09}). Due to the finite size of the considered system the actual percolation threshold is not sharp, and tortuosity around $\varphi_c$ is underestimated in this area. The origin of the drop in the tortuosity below $\varphi_c$ is the finite size of samples. 

To summarize, we have shown that the convolutional neural networks can be used to predict the fundamental quantities of the porous media: porosity, permeability and, tortuosity based on the pictures representing the two-dimensional systems.  
The two types of the networks have been discussed. In the first, the input pictures were resized, in the other the network took the original size input. We obtained good accuracy predictions of all three quantities. The networks reconstruct the empirical dependence between tortuosity and porosity obtained in the previous studies\cite{Matyka08}. 

\bibliography{references}

\section*{Author contributions statement}
%Must include all authors, identified by initials, for example:
%A.A. conceived the experiment(s),  A.A. and B.A. conducted the experiment(s), C.A. and D.A. analysed the results.  
K.G. and M.M. conceived and conducted the project. KG designed and performed the deep learning analysis. MM designed and performed fluid flow simulations. K.G. and M.M. wrote and reviewed the manuscript.

\section*{Additional information}
%To include, in this order: 
%
%\textbf{Accession codes} (where applicable); 

%\textbf{Competing interests} (mandatory statement). 
\textbf{Competing Interests:} The authors declare that they have no competing interests.

%The corresponding author is responsible for submitting a \href{http://www.nature.com/srep/policies/index.html#competing}{competing interests statement} on behalf of all authors of the %paper. This statement must be included in the submitted article file.

\end{document}